\documentstyle[12pt,lnfprep]{article}
\newcommand{\Header}{
  \begin{tabular}{rl}
  \hspace{-.4cm}\includegraphics{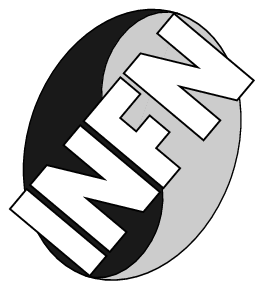} &
    \renewcommand{\arraystretch}{0.5}
    \begin{tabular}{r}
      {\hspace{1cm}~\LARGE\sffamily LABORATORI~ NAZIONALI~ DI~ FRASCATI}\\
      \\
      {\Large\sffamily SIS-Pubblicazioni}\\
    \end{tabular}
    \renewcommand{\arraystretch}{1}
  \end{tabular}
  \vskip 1cm
  \begin{flushright}
  \renewcommand{\arraystretch}{0.5}
    \begin{tabular}{r}
      {\underline{LNF-97/040(P)}}\\    
      \\
      {\small 26 Novembre 1997}       
    \end{tabular}
  \end{flushright}
  \renewcommand{\arraystretch}{1}
  \vspace{2cm} 
  }
\def\DAFNE{${\rm DA}\Phi {\rm NE}$ }
\def\FI{$\phi$}
\def\ee{$\,e^+e^-\,$}
\def\cms{ {\rm cm}^{-2} {\rm s}^{-1} }
\def\dedx{ $-(dE/dx)$ }
\begin{document}
\begin{titlepage}
\title{ 
  \Header
  {\large \bf 
DEAR, FINUDA, KLOE: \\
Kaonic Atoms, Hypernuclei and CP-Violation \\
at the DA$\Phi$NE $\Phi$-Factory
}
}
\author{
   Stefano Bianco\\
   {\it INFN-Laboratori Nazionali di Frascati,} \\
   {\it Via Enrico Fermi 40, 00044 Frascati, Italy}
}
\maketitle
\baselineskip=14pt

\begin{abstract}
Physics at \DAFNE, the new Frascati \ee machine, is reviewed, as well
as the experiments:  DEAR - search for $KN$ exotic atoms,
FINUDA - spectroscopy
and decays of hypernuclei,  and KLOE - a  multipurpose detector designed
for detecting direct CP violation.
\end{abstract}

\vspace*{\stretch{2}}
\begin{flushleft}
\end{flushleft}
\begin{center}

Invited Talk Delivered at the II SILAFAE \\
November 1996, Merida, Yucatan (Mexico). 
\end{center}
\end{titlepage}
\pagestyle{plain}
\baselineskip=17pt
\section{Introduction}
 The \DAFNE  \ee collider facility
\cite{maiani95},\cite{dafne95},\cite{weblnf}, proposed in
1990 \cite{yellow90}, 
has been completed in mid 1997 at the Laboratori Nazionali di
Frascati of INFN, and it is now being commissioned.
The \DAFNE collider\cite{epac94}\cite{vignola95} will operate at
the center of mass energy
of the $\phi$ meson with an initial luminosity ${\cal L} = 1.3\times 10^{32}
\cms$  and a target luminosity ${\cal L} = 5.2\times 10^{32} \cms$.
\par
The \FI(1020) meson  
\footnote{ Quantum numbers and physical constants for the \FI(1020) meson
are:  $s{\bar s}$ quark assignment,
mass $m=1019.413 \pm 0.008\, {\rm MeV}$, total  width
$\Gamma = 4.43 \pm 0.06 \,{\rm MeV}$,
electronic width $\Gamma_{ee} = 1.37 \pm 0.05 \, {\rm keV}$,
and $I^G(J^{PC})=0^-(1^{--})$  quantum numbers.}
is produced in \ee collisions with a
$$\sigma(e^+e^- \rightarrow \phi) =  4.4\,\mu {\rm b}$$
peak cross section, which translates into a production rate of
$$R_{\phi}= 2.3 \times 10^3 {\rm s}^{-1}$$
at target luminosity.
The production cross section $\sigma(e^+e^- \rightarrow \phi)$
should be compared
with the  hadronic production which, in this energy range,
is given by
$$\sigma(e^+e^- \rightarrow f {\bar f}) = \frac{4\pi \alpha^2}{3s} Q_f^2
              = \frac{86.8 \, Q_f^2 \,{\rm nb}}{s[{\rm GeV}^2]}
              = 56\,{\rm nb},$$
i.e., with an integral S/N ratio of about 40:1.
\par
The \FI~ then decays at rest into
$K^+K^-$ (49\%),
$K^0 \bar{K^0}$ (34\%),
$\rho \pi$ (13\%),
$\pi^+ \pi^- \pi^0$ (2.5\%)
and $\eta \gamma$ (1.3\%),
which translate in
0.8~kHz $\phi\rightarrow K^0 \bar{K^0}$,
1.1~kHz $\phi\rightarrow K^+ K^-$,
300~Hz $\phi\rightarrow \rho \pi$,
60~Hz $\phi\rightarrow \pi^+\pi^-\pi^0$,
30~Hz $\phi\rightarrow \eta\gamma$.
With a canonical efficiency of 30\%, this corresponds to
$2\times 10^{10}$ $\phi$ decays in one calendar year.
Therefore, a \FI-factory is a unique source of
monochromatic (110 and 127 MeV/c, respectively),
slow
\footnote{ Kaons are produced with $\beta_K \simeq 0.2$.
Therefore they travel a short path before decaying, and can be stopped
after crossing very little matter. Experimentally, this translates into
small detectors and thin targets.},
 collinear,
quantum-defined
(pure $J^{PC}=1^{--}$ quantum states),
and tagged
\footnote{The detection of  one K out of the two produced in the \FI~ decay
determines the  existence and direction of the other K.}
neutral and charged kaons.
\par
With
${\cal O} (10^{10})$ kaons per year we can
form kaonic atoms  and study K-Nucleon interactions
at low energy (DEAR), produce nuclei with strangeness ({\it hypernuclei},
FINUDA), and study particle physics fundamental symmetries (KLOE).
\section{A brief history of kaons }
The kaonic enigma \cite{perkins},\cite{predazzi}
began in 1947 with their very discovery
in p-N interactions such as $\pi^- + p \rightarrow \Lambda + K^0$.
The long lifetime $\sim 10^{-10} {\rm s}$, and the fact they were
always produced in pairs, was explained by postulating
a new quantum number {\it strangeness} conserved by strong interactions
(production) but not weak interactions (decay) (Tab.\ref{tabkaons}).
\begin{table}[t]
\centering
\caption{ Quantum number assignments for kaons.
}
\vskip 0.1 in
\begin{tabular}{|l|r|r|} \hline
       &   $I_3=+1/2$            & $I_3=-1/2$         \\
\hline
\hline
$S=+1$ & $K^+ (\bar{s} u)$       & $K^0 (\bar{s} d)$  \\
$S=-1$ & $\bar{K^0} (s \bar{d})$ & $K^- (s \bar{u})$  \\
\hline
\end{tabular}
\label{tabkaons}
\end{table}
In 1956, Lee and Yang postulated first, and Wu et al.
demonstrated next (1957), that the $\tau-\theta$ puzzle [the same $K^+$
particle decaying to opposite parity $(2\pi, 3\pi)$ states]
was actually caused by P and C being violated in weak decays,
with CP  conserved. In 1964   Christenson, Cronin, Fitch and
Turlay disproved this hypothesis by studying $K^0$ decays.
$K^0$ and $\bar{K^0}$ have same quantum numbers except S, which is not
conserved in weak interactions; therefore, they have common  virtual  decay
channels and they can oscillate when decaying
\footnote{The $3\pi$ state has
  $CP=-1$  when in the L=0 state, with 
the L=1 state being depressed by centrifugal
potential.}:
$$K^0 \leftrightarrow 2\pi \,  (CP=+1) \leftrightarrow \bar{K^0} \qquad
K^0 \leftrightarrow 3\pi \,  (CP=-1) \leftrightarrow \bar{K^0}$$
with the CP eigenstates being
$$|K_{1,2}^0\rangle  = (|K^0\rangle  \pm |\bar{K^0}\rangle )/ \sqrt 2$$
and time evolution
$$P(K^0,t)=1/4 [ e^{-\Gamma_1t}+
e^{-\Gamma_2t}-2e^{-(\Gamma_1+\Gamma_2)t/2}
\cos\Delta m t].$$
The decay of 
neutral kaons to pions is Cabibbo-unfavoured: since the
phase space available for  $2\pi$ is much greater than for  $3\pi$, lifetimes
will be largely different $\Gamma_1 >> \Gamma_2$.
Quite fortuitously,
the oscillation frequency is
$\Delta m \equiv |m_1-m_2|$, large and observable,
since the mass difference is comparable to the width of the short-lived state
${ {\Delta m}\over {\Gamma_1} } \sim 0.5.$
Christenson et al. indeed showed  that  a
$10^{-3}$ fraction of $K_2$'s decay to $2\pi$.
Physical states are therefore
$$|K_L\rangle  \propto           |K_2\rangle  + \epsilon |K_1\rangle \quad
{\rm and} \quad |K_S\rangle \propto \epsilon |K_2\rangle  +  |K_1\rangle.$$
The $\epsilon$ parameter describes {\it indirect} CP violation (CPV in short),
which originates from the mass matrix mixing
($\Delta S=2$ transitions), not from the interaction responsible
of the decay.
\par
The central role played by the $K^0-\bar{K^0}$ mixing was revived in
1970. The suppressed nature of neutral-current decays
 such as $K^0_L\rightarrow \mu\mu$ was
explained through GIM-like box-diagrams
and predicted a fourth flavour -
{\it charm} - with mass $m_C^2 \propto \Delta m$.
\par
Together with CPV from mixing, there can be a {\it direct}
$(\Delta S=1)$
transition from $|K_L (CP=-1)\rangle$ to $|2\pi (CP=+1)\rangle$, without
$K^0 \leftrightarrow \bar {K^0}$
intermediate transition. Direct CPV is parametrized by
$$\epsilon^\prime \propto
{ {\langle 2\pi \, I=2 | H | K^0\rangle} \over
{\langle 2\pi\, I=0 | H | K^0\rangle} }. $$
\par
The issue of CPV is a most fundamental one. As first emphasized by
Sakharov in 1967, the predominant asymmetry
of baryons versus antibaryons in the universe
could be explained by three requirements:
baryon number non-conservation, a CPV baryon-creating process,
 with baryons out of the thermal equilibrium. See M.~Gleiser's
talk at this Symposium \cite{gleiser}.
\par
Models proposed so far can be schematically distinguished in two
classes, milliweak and superweak.
Milliweak models advocate that a tiny $(10^{-3})$ fraction
of  the weak interaction is indeed CPV at
leading order $(\Delta S=1)$. They imply T violation to guarantee
the CPT theorem, and they can explain both  direct [one
$(\Delta S=1)$ transition] and indirect [two $(\Delta S=1)$
transitions]. Superweak models (Wolfenstein, 1964), on the other hand,
postulate a new $(\Delta S=2)$,  CPV process which
transforms $K_L\rightarrow K_S$. The strength
of this interaction relative to weak coupling is  only $10^{-10}$, since
$M_{K_S} \simeq M_{K_L}$ and,  therefore, no direct CPV is
predicted nor allowed, no CPV is expected in other systems.
The standard model (SM), a milliweak-type theory,
{\it naturally} allows for CPV: the angle $\delta$ in
the CKM matrix allows for CPV transitions
\begin{equation}
V_{CKM}=
\left(
\begin{array}{ccc}
      1-\lambda^2/2     &   \lambda      &   A\lambda^3\sigma e^{-i\delta} \\
      \lambda           & 1-\lambda^2/2  &       A\lambda^2                \\
      A\lambda^3(1-\sigma e^{i\delta})    & -A\lambda^2 & 1
\end{array}
\right)
+
{\cal O} (\lambda^3)
\end{equation}
where $\lambda \equiv \sin \theta_C$,
$\rho - i\eta \equiv \sigma e^{-i\delta}.$
\par
\begin{figure}[t]
\vspace{3cm}
\includegraphics{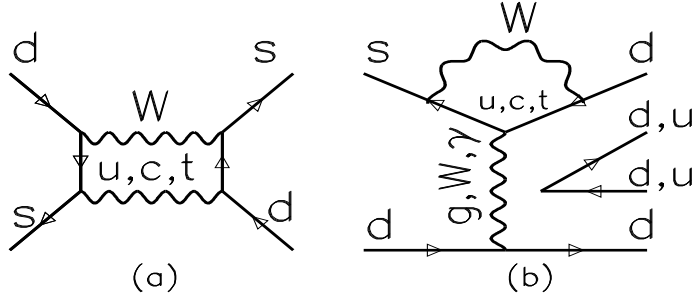}
\caption{ a) $K^0-\bar{K^0}$ mixing b) Direct CP violation through
{\it penguin} diagrams.}
\label{diagrams}
\end{figure}

The indirect CPV parameter $\epsilon$ is proportional
in the SM picture to $\Delta S=2$ box diagrams (Fig.\ref{diagrams}a)
$\epsilon \propto \lambda^4 \sin \delta \simeq 10^{-3} \sin \delta$.
Therefore, $\epsilon$ is small per se because of the suppression of
interfamily transitions, while the CPV angle $\delta$ is not
necessarily small. Experimentally, $|\epsilon| = 2.26(2) \times 10^{-3}.$
\par
The direct CPV parameter $\epsilon^\prime$, on the other hand,
is proportional in the SM picture to $\Delta S=1$ {\it penguin} diagrams
(Fig.\ref{diagrams}b) which are difficult to compute,
and suppressed by both the $\Delta I=1/2$ and the Zweig rule
\footnote{We do not discuss here the third piece of experimental
  information
that can be used to set limits to the CPV angle $\delta$,
i.e. the experimental bounds on the electric dipole moments of the
neutron and the electron.
See \cite{peccei95} and references therein for a thorough review.}.
Calculations\cite{buchalla91} have shown how in the SM picture
direct CPV is a decreasing function of the quark masses in the penguin
loop, primarily the top quark mass. Prediction for $m_{top}=175\,{\rm GeV}$
is $0\leq \epsilon^\prime/\epsilon \leq 0.001$.
\par
\begin{figure}[t]
\vspace{3.5cm}
\includegraphics{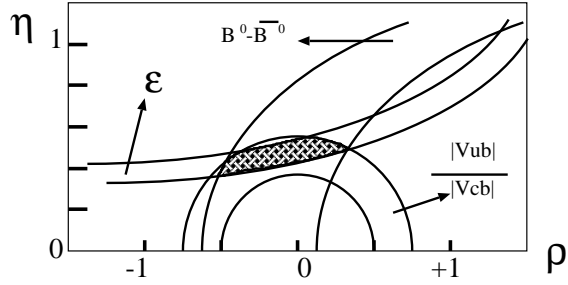}
\caption{ Qualitative sketch showing bounds on the CP violating
angle of CKM matrix. }
\label{constraints}
\end{figure}

One can try and use the precise measurement of $|\epsilon|$ to set bounds
on $\delta$, although along with the use of the experimental values from the
$B_d^0-\bar{B}_d^0$ mixing and the ratio of CKM elements
$|V_{ub}|/|V_{cb}|$, obtaining \cite{peccei95} the qualitative picture in
Fig.\ref{constraints}.
\par
Finally, the most precise measurements
\footnote{
Experimentally,   $|\eta_{+-}|\simeq |\eta_{00}|\simeq
2\times 10^{-3}$ and $\phi_{+-}\simeq\phi_{00}\simeq 44^o$.
Therefore, $\epsilon^\prime/\epsilon \simeq \Re
(\epsilon^\prime/\epsilon)$.
}
 of the relative strength of direct versus indirect CPV yield result in
mild agreement:
$\Re (\epsilon^\prime / \epsilon) = (7.4\pm 5.2 \pm 2.9) \times
10^{-4}$ (E731 at Fermilab\cite{e731}) and
$\Re (\epsilon^\prime / \epsilon) = (23.0\pm 3.6 \pm 5.4) \times
10^{-4}$ (NA31 at CERN\cite{na31}) as measured  from the double-ratio
\footnote{The following useful relationships also hold:
$$\eta_{+-} \equiv \epsilon + \epsilon^\prime \equiv
{ {\langle \pi^+\pi^- | H | K_L\rangle} \over
{\langle \pi^+\pi^-| H | K_S\rangle} }     \quad
\eta_{00} \equiv \epsilon - 2 \epsilon^\prime \equiv
{ {\langle \pi^0\pi^0 | H | K_L\rangle} \over
{\langle \pi^0\pi^0| H | K_S\rangle} }. $$
}
\begin{equation}
\Big| \frac{\eta_{+-}}{\eta_{00}} \Big|^2 =
{ {N(K_L \rightarrow \pi^+\pi^-) / N(K_S \rightarrow \pi^+\pi^-) }
\over
    {N(K_L \rightarrow \pi^0\pi^0) / N(K_S \rightarrow \pi^0\pi^0) } }
\simeq
1+6 \Re (\epsilon^\prime / \epsilon)
\label{eq:doubleratio}
\end{equation}
\par
From the body of information available, the CPV angle $\delta$ seems
to be large $(\rho \simeq 0 \Rightarrow \delta\simeq \pi/2)$; a large top
mass (175 GeV) predicts $\Re (\epsilon^\prime/\epsilon)$ to be in the
$10^{-4}$ region \cite{buchalla91}; a  measurement of
$\Re (\epsilon^\prime/\epsilon) \not=0$ will indicate the general
validity of the CKM picture, since it requires the presence of a
$\Delta S=1$ phase.
\par
\par
\section{KLOE}
The KLOE experiment
was proposed in April 1992; approved and funded at the
beginning of 1993, construction of the detector began in 1994.
The principal aim  of KLOE is the detection of direct CP violation in $K^0$
decays with a $10^{-4}$ sensitivity on $\Re (\epsilon'/\epsilon)$. For
a comprehensive review see \cite{berto94} and references therein, while an
up-to-date  status report can be found in \cite{juliet96}.
The measurement in KLOE of $\Re (\epsilon'/\epsilon)$ with the
required precision takes advantage of unique features of a $\phi$-factory:
numerous kinematical constraints exist, $K_S$ and $K_L$ are detected
at the same time, clean environment typical of \ee machines with respect to
experiments at extracted hadron beams, built-in calibration processes.
On the down side, the $K_S$ and $K_L$ are detected  with different
topologies
 in both the charged and the neutral decay modes.
\par
KLOE
should be regarded as a precision kaon interferometer.
As already stressed, collinear
kaon pairs are produced from the $\phi$ decay in a pure, coherent quantum
state.
If one of the two neutral kaons from the $\phi$ decays into a state $f_1$
at time $t_1$ and the other to a state $f_2$ at time $t_2$, the decay
intensity to $f_1$ and $f_2$ as a function of $\Delta t \equiv t_1-t_2$ is
\begin{eqnarray}
I(f_1,f_2; \Delta t ; \forall \Delta t > 0) &   =   &
\frac{1}{2\Gamma} | \langle f_1| K_S \rangle \langle f_2|K_S\rangle|^2
\big[ |\eta_1|^2 e^{-\Gamma_L\Delta t} +
      |\eta_2|^2 e^{-\Gamma_S\Delta t} -        \nonumber \\
 &    &
      2||\eta_1||\eta_2|e^{-\Gamma \Delta t/2}
           \cos (\Delta m \Delta t + \phi_1 -\phi_2)      \big]
\end{eqnarray}
where $\eta_i \equiv \langle f_i|K_L\rangle / \langle f_i|K_S\rangle$,
and $\Gamma \equiv (\Gamma_S+\Gamma_L)/2$.
The interference term in the
decay intensities above gives measurements of all sixteen parameters
describing CP and CPT (if any) violations in the neutral kaon
system \cite{peccei95}. As instance, when $f_1=f_2$, one measures $\Gamma_S,
\Gamma_L$ and $\Delta m$; when $f_1=\pi^+\pi^-$, $f_2=\pi^0\pi^0$, one
measures $\Re (\epsilon'/\epsilon)$ at large $\Delta t$, and
 $\Im (\epsilon'/\epsilon)$ near $\Delta t=0$.
\par
Other physics topics include rare $K_S^0$ decays ($10^{10}$ kaons per year
will improve the sensitivity to branching ratios down to the $10^{-8}$ range),
tests of Chiral Perturbation Theories, radiative  \FI~ decays, and
$\gamma \gamma$ physics \cite{anulli95}.
Out of such a physics spectrum other than CPV, it should be underlined the
importance of a study of the $J^{PC}=0^{++}$  $f_0(975)$ meson state,
produced via the $\phi$ radiative decay with a branching ratio
at least $1\times 10^{-6}$. The nature of the lightest scalar is still
unclear, whether a $KK$ molecule, an exotic $qq\bar{q}\bar{q}$ or a glue-ball
state. The decay $f_0\rightarrow \pi^0\pi^0$ should be easily measured
even at low luminosity during the machine commissioning
\cite{franzini92}, while the charged channel $\pi^+\pi^-$ suffers
by backgrounds from continuum.
\par
To reach the design sensitivity on the measurement
of $\Re (\epsilon'/\epsilon)$ via the classical double-ratio method
(Eq.\ref{eq:doubleratio}), a one-year
statistics at full \DAFNE luminosity is necessary. It is also necessary to
control the detector efficiency for the decays of interest and to reject
backgrounds from the copious $K_L^0$ decays to states other than two-pion.
\par
\begin{figure}[t]
\vspace{8cm}
\includegraphics{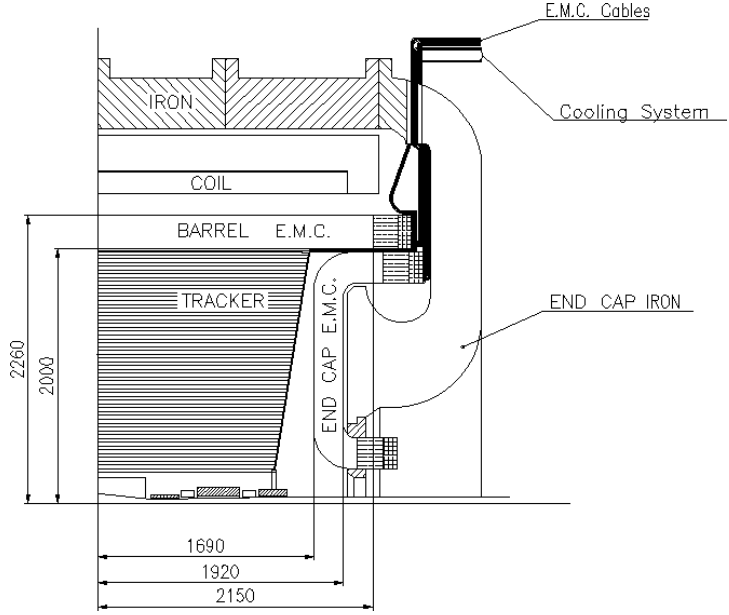}
\caption{ Cross-sectional schematics of the KLOE detector (one quadrant, side
view). Dimensions are millimeters.}
\label{klomerida96}
\end{figure}

The KLOE detector \cite{kloedeteref}
(Fig.\ref{klomerida96}) is a hermetic, 4$\pi$ geometry
apparatus: a cylindrical structure surrounding the beam pipe and consisting of
a large drift chamber, and  an electromagnetic calorimeter with
state-of-the-art energy and time resolution
\cite{hopsnim},\cite{anto95},\cite{misc96}, and  a 6~kG superconducting magnet.
Counting of the four $K_S, K_L$ decays used in the double-ratio formula
(eq.\ref{eq:doubleratio}) is performed by defining a $K_S$ fiducial volume
(10~cm radius around the interaction point, about 17 lifetimes), and a $K_L$
fiducial volume (145~cm, about one half  lifetime).
It is very important to avoid systematical errors in the determination
of the boundaries of the fiducial volumes for neutral and charged decay
channels. This is accomplished by using $K_L$ decays where both the neutral
and the charged vertex are measured, as in $K_L\rightarrow \pi^+\pi^-\pi^0$.
\par
\par
The KLOE calorimeter will reconstruct the $\pi^0\pi^0$ mode,
determine the decay vertex space location, reject the $3\pi^0$ decay, and
provide $\pi/\mu$ rejection. The technique employed is a Pb-scintillator
sampling with 1-mm scintillating fibers embedded in very thin Pb grooved foils,
which has been demonstrated to provide good energy
$(\sigma_E/E \simeq 4\%/\sqrt{E [{\rm GeV}]})$
and   space ($\sigma_{x,y}\simeq 0.5\,{\rm cm}/\sqrt{E}$,
$\sigma_z \simeq 2\, {\rm cm}$) resolution,
 acceptable efficiency
for photon energies down to 20~MeV, and spectacular time resolution
$(\sigma_t = 55\,{\rm ps} /\sqrt{E[{\rm GeV}]})$.
The measurement of the time of arrival $t_\gamma$ and the
impact point $(x,y,z)$ at the calorimeter 
of one photon out of four,
and the knowledge of the $K_S$ flight direction $(\cos\theta)$,
gives the $K_L$ decay length $L_K$ to an accuracy of
1~{\rm cm} via the formulae
\begin{equation}
t_\gamma = L_K/(\beta_K c) + L_\gamma/c 
\end{equation}
\begin{equation}
L_\gamma^2 = D^2 + L_K^2 -2DL_K \cos\theta
\end{equation}
where $D^2\equiv x^2+y^2+z^2$.
\par
A He-based, very transparent, drift chamber will reconstruct the $\pi^+\pi^-$
final state, reject the $K_{\ell 3}$ background, and
determine the $K_S^0$ flight
direction and the charged decay vertices.
The KLOE drift chamber has space resolution
$\sigma_{\rho,\phi}=200\,\mu {\rm m}$ and $\sigma_z=3\,{\rm mm}$, and a
0.5\% relative
resolution on the transverse momentum.
\par
To achieve the required statistical sensitivity, the entire \FI~ event rate
(5~kHz) has to be written on tape. The calorimeter will provide triggering for
most of the decay channels of interest.
The rejection of three neutral-pion decays relies on the calorimeter's
hermiticity and efficiency for photons down to 20~MeV energy. The rejection of
the $K_{\mu3}$ decay over the $\pi^+\pi^-$ decay of interest is based on several
kinematical variables and constraints. The high-resolution features of the
tracking limits the residual contamination to $4.5 \times 10^{-4}$,
at a 6~kG solenoidal field, with a 99.8\% $\pi^+\pi^-$ efficiency.
\par
KLOE expects to be operational at beginning of 1998 \cite{kloe97}.
\section{FINUDA}
Nuclear physics topics will be investigated by the FINUDA (standing for {\it
FIsica NUcleare a DA$\Phi$ne}) experiment.
 A nuclear physics experiment carried
out at an \ee collider sounds  contradictory in itself, but this is
where the
uniqueness of the idea lies \cite{bressani91}: charged  kaons from
\FI~ decays are used as a monochromatic, slow (127~MeV/c), tagged,
background-free, high-counting rate beam on a thin target surrounding the beam
pipe. The possibility of stopping low-momentum monochromatic $K^-$ with a thin
target
(typically $0.5\, {\rm g} \, {\rm cm}^{-2}$ of
$^{12}C$) is unique to  \DAFNE: $K^-$'s can
be stopped with minimal straggling very near the target surface,
so that outgoing
prompt pions do not cross any significant amount of the
target and do not undergo
any momentum degradation. This feature
provides unprecedented momentum resolution as long as very transparent
detectors are employed before and after the target.
\par
The FINUDA detector is optimized to perform high-resolution studies of
hypernucleus production and non-mesonic decays\cite{bressani95},
by means of a spectrometer
with the large acceptance typical of collider experiments.
\par
Negative kaons stopping inside the
target produce a $Y$-hypernucleus $(Y=\Lambda,\Sigma,...)$ via the process
\begin{equation}
K^-_{stop} + {~^A{Z}} \rightarrow \, {~^{A}_Y{Z}} + \pi^-,
\label{eq:hypernucleus}
\end{equation}
where the momentum of the outgoing $\pi^-$ is directly
related to the level of
the hypernucleus formed (two-body reaction).
In the case of $\Lambda$ hypernucleus formation, the
following weak-interaction 'decays'
are strongly favored in medium-heavy  nuclei
$$\Lambda + n \rightarrow n + n \qquad\qquad \Lambda + p \rightarrow n + p,$$
with the nucleus undergoing the reactions
$${~_{\Lambda}^A{Z}} \rightarrow ~^{(A-2)}Z     + n + n \qquad\qquad
{~_{\Lambda}^A{Z}} \rightarrow ~^{(A-2)}(Z-1) + n + p,$$
which are interesting for studying the validity of the $\Delta I=1/2$
rule \cite{bressani95}.
\par
Hypernuclei are a unique playground for both nuclear and particle
physics \cite{fabbri96}. A $\Lambda$ embedded in a nucleus can occupy,
because of the
strangeness content, levels forbidden to a $p$ and a $n$ by the Pauli
exclusion principle. Furthermore, the $I=0$ assignment makes the
$\Lambda N$  interaction much weaker than the ordinary $NN$ interaction.
 In the shell model of nuclei the quarks are confined in bags inside
 baryons.  Baryons inside nuclei maintain their identity and interact by
 the exchange of mesons. At short distance the bags can fuse, quarks can
 get deconfined and begin interacting by exchanging gluons. For a
 $~_{\Lambda}^5{He}$ hypernucleus the two models give different
 predictions: while in the baryon model the $\Lambda$ occupies the
 $S_{\frac{1}{2}}$ orbital due to its strangeness quantum number, in the
 quark model the $u$ and $d$ quarks inside the $\Lambda$ cannot stay in the
 $S_{\frac{1}{2}}$  but are obliged to move to the  $P_{\frac{3}{2}}$
orbital by Pauli blocking. As a result, different binding energies are
predicted by the models.
\par 
Besides studies on hypernucleus spectroscopy, interest exists for
K-N interactions at low energy \cite{olin95}.
Recently,
the possibility of improving the precision on the $K_{e2}$ branching ratio
by a factor of 3 over the present world average was also discussed
\cite{bressanike2}.
\par
\begin{figure}[p]
\vspace{18cm}
\includegraphics{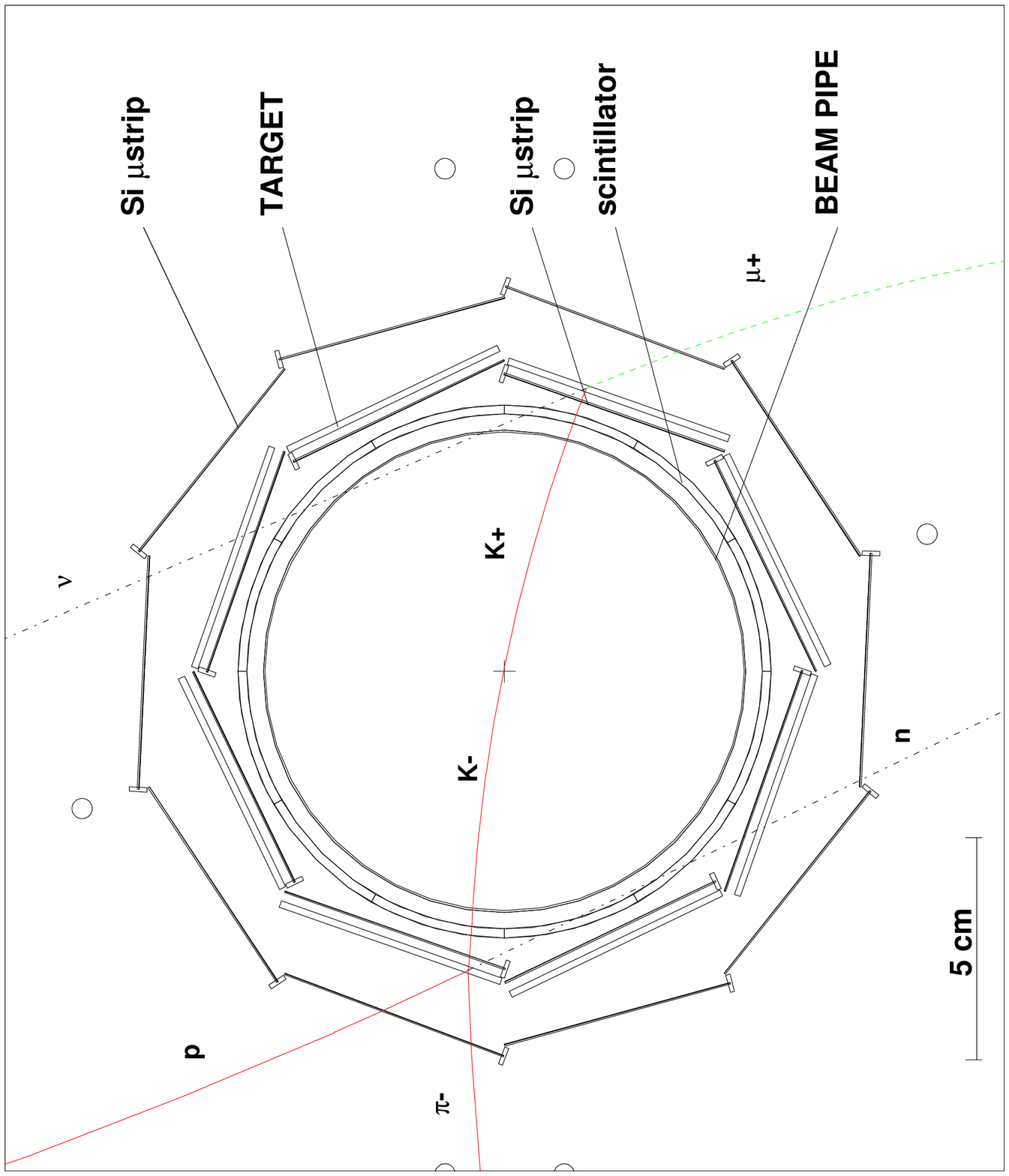}
\includegraphics{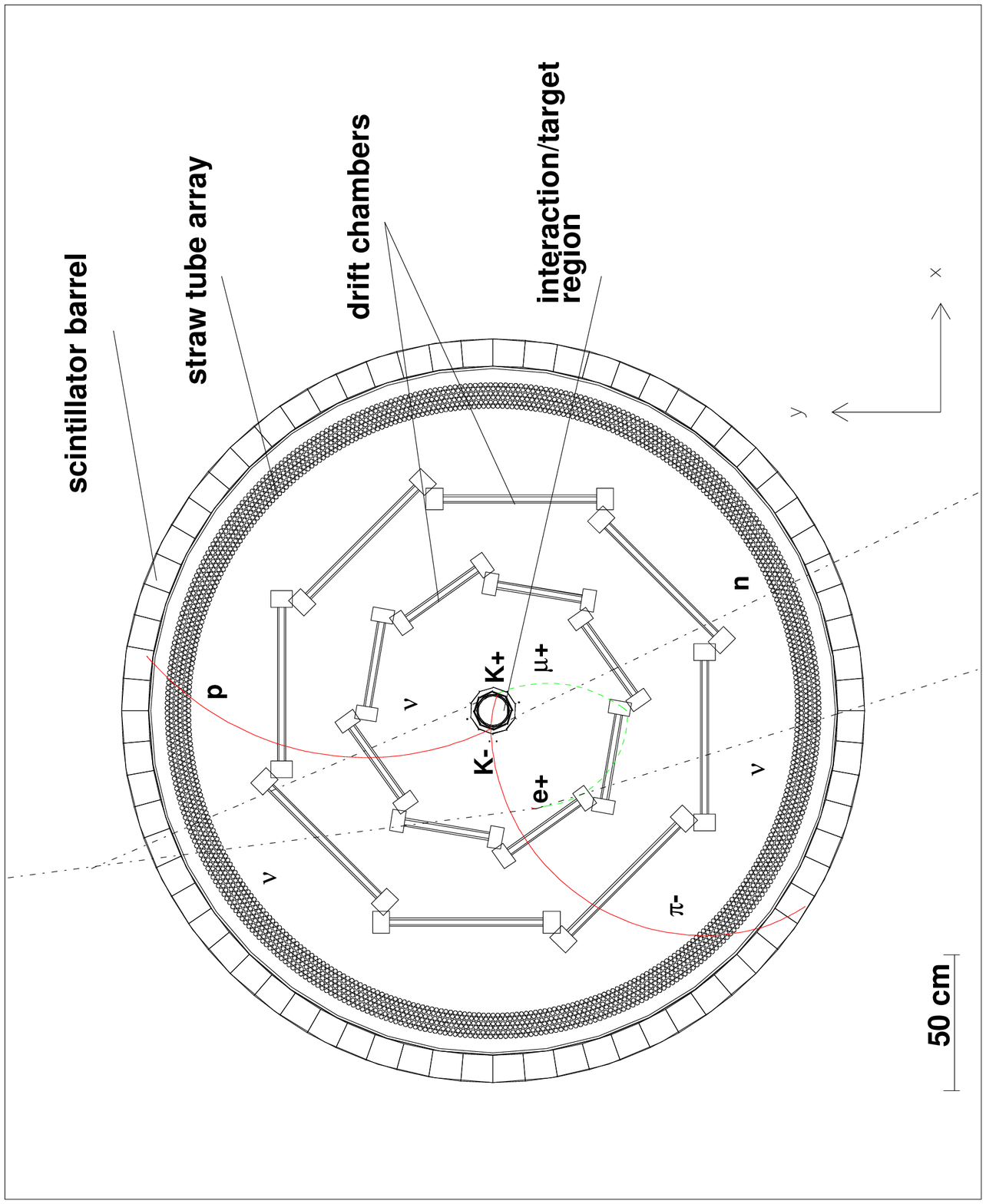}
\caption{
Sketch of the FINUDA detector front (beam) view.
The interaction-target region (top); the outer tracker and scintillator
barrel (bottom). An hypernuclear event is superimposed.
A $\phi$ decays to $K^+K^-$. The $K^+$ decays to $\mu^+ \nu_\mu$, followed
by $\mu^+\rightarrow e^+ \nu_e \bar{ \nu}_\mu$. The $K^-$
crosses the inner trigger scintillator, is tracked by the inner
Si-$\mu$strip detector, stops in the target and forms
an hypernucleus (eq.\ref{eq:hypernucleus}).
The momentum of the emitted prompt $\pi^-$,
proportional to the energy of the hypernuclear level formed, is
measured by the spectrometer. The
hypernuclear $\Lambda$  undergoes the process $\Lambda \,p \rightarrow
n\,p$.
}  
\label{detec}
\end{figure}

FINUDA  \cite{finudaprop}
is a magnetic spectrometer with
cylindrical geometry,
optimized in order to have large solid-angle acceptance,
optimal momentum resolution
(of the order of $ 0.3 \, \%$ FWHM on prompt pions)
and good triggering capabilities. The geometrical acceptance
is 100\% in the $\phi$ angle (which reduces to 80\% due
to dead material), and
     approximately $135^0 \leq \vartheta \leq 45^0$, thus naturally
rejecting  \ee background from Bhabha scattering.
FINUDA consists of (Fig.\ref{detec}) an interaction/target region,
external tracker, outer scintillator array, and superconducting solenoid.
\par
\par
The ($K^+,K^-$) pairs from \FI~ decay emerge from the interaction
region (Fig.\ref{detec}), and are identified for triggering by
exploiting the back-to-back event topology and  by identification
through  \dedx in the TOFINO triggering
scintillator array.
Before impinging on the target, a Si-$\mu$strip array
 measure the ($K^+,K^-$) coordinates before the nuclear target
with $\sigma \sim 30-50\mu {\rm m}$ and provides particle identification
by \dedx. The monochromatic $K^-$ is stopped inside the thin target,
reaction products are emitted isotropically: the momentum of the prompt
pion
is proportional to the energy of the hypernuclear level formed, and it is
measured by Si-$\mu$strip, low-mass drift chambers\cite{agnello95},
and  straw tube arrays \cite{benussi94}. The signal in the outer
scintillator array (TOFONE) provides  a fast trigger logic coincidence.
Baryons or mesons from $\Lambda$ decays are also tracked in the
spectrometer (protons, $\pi ^\pm$) or detected in the TOFONE (neutrons).
The expected FWHM  resolution on the  270 MeV/c prompt pion is
0.25\% for forward pions (i.e., those emitted towards the outer region
of the spectrometer), which corresponds
to a 0.7~MeV resolution on the hypernuclear level,
to be compared to the best result achieved so
far at fixed-target of about 2~MeV \cite{hasegawa95}.
\par
The rate of reconstructed hypernuclear events with resolution better
than 1~MeV at ${\cal L} = 10^{32} {\rm cm}^{-2} {\rm s}^{-1}$
is given by the expression:
$$
R({_\Lambda}Z) = R_\phi                             \times
                 B_{K^+K^-}        \times
                 {{N_{{_\Lambda}Z}}\over{K_{stop}}} \times
                 \epsilon_{\pi^-}                   \times
                 \epsilon_{transp}
                 \simeq 2.8 \times 10^{-2} {\rm s}^{-1}
                 \simeq 100\,{\rm evnts/hr}   $$
where $R_\phi= 5\times 10^{2} {\rm s}^{-1}$,
$B(\phi \rightarrow K^+K^-)=0.49$,
$N_{{_\Lambda}Z}/K_{stop}=10^{-3}$ is the capture rate,
$\epsilon_{\pi^-}=13\%$ is the total efficiency for forward
(high-resolution) prompt pions, and $\epsilon_{transp}=80\%$ is the
chamber transparency.
\par
\begin{figure}[p]
\vspace{20cm}
\includegraphics{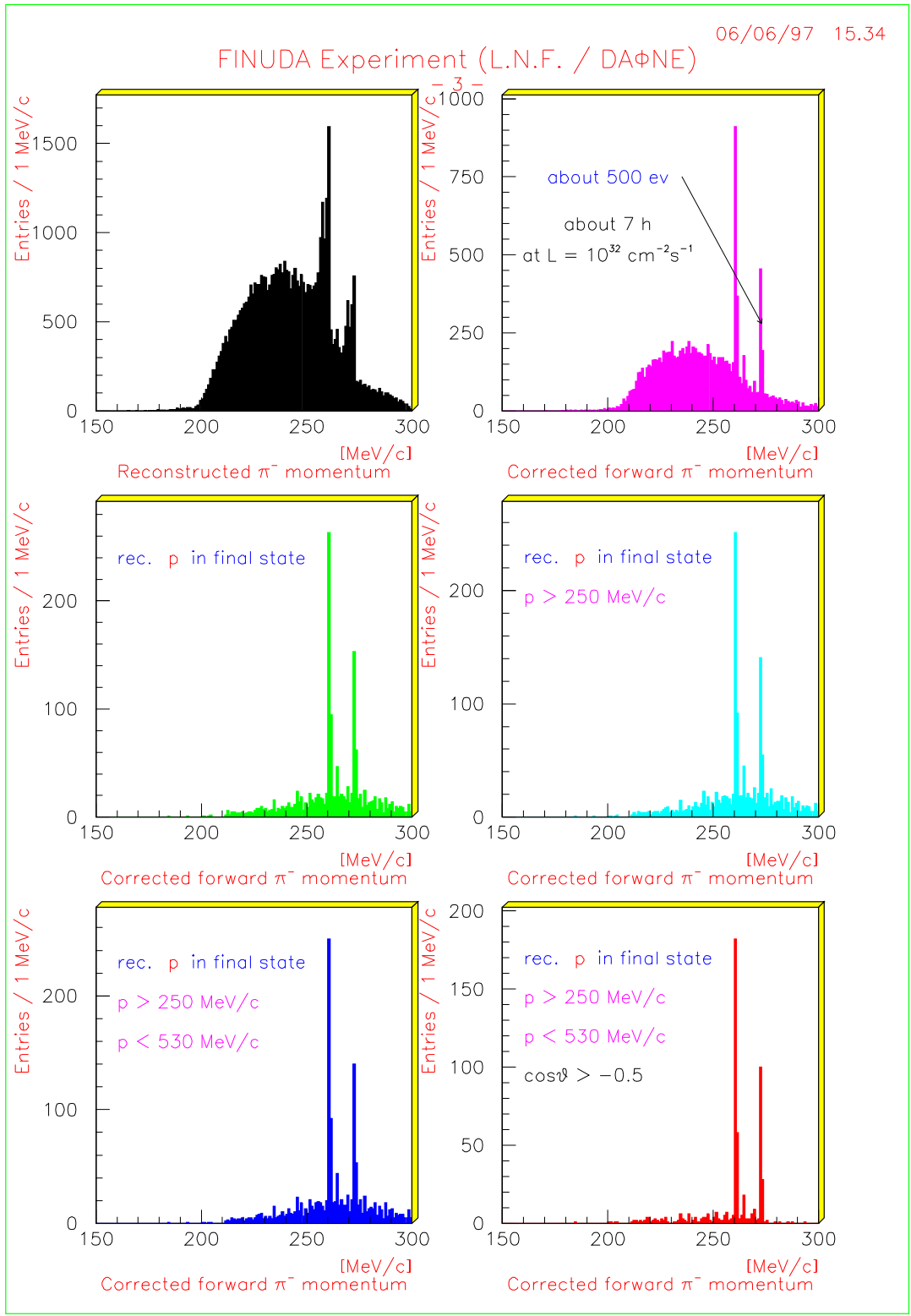}
\caption{The effect of increasingly stringent cuts on a four-state
simulated hypernuclear spectrum, as reconstructed by the
FINUDA spectrometer.}
\label{finpeaks}
\end{figure}

Superior momentum resolution, high statistics, and low background,
make FINUDA capable of resolving hypernuclear states produced
at a relative rate of order 
${{N_{{_\Lambda}Z}}\over{K_{stop}}} = 10^{-4}$ as simulated
in Fig.\ref{finpeaks} (from ref.\cite{zenoni95}).
\par
The FINUDA experiment expects to be operational  by beginning of 1998.
\par
\par
\section{DEAR}
\begin{figure}[pt]
\vspace{10cm}
\includegraphics{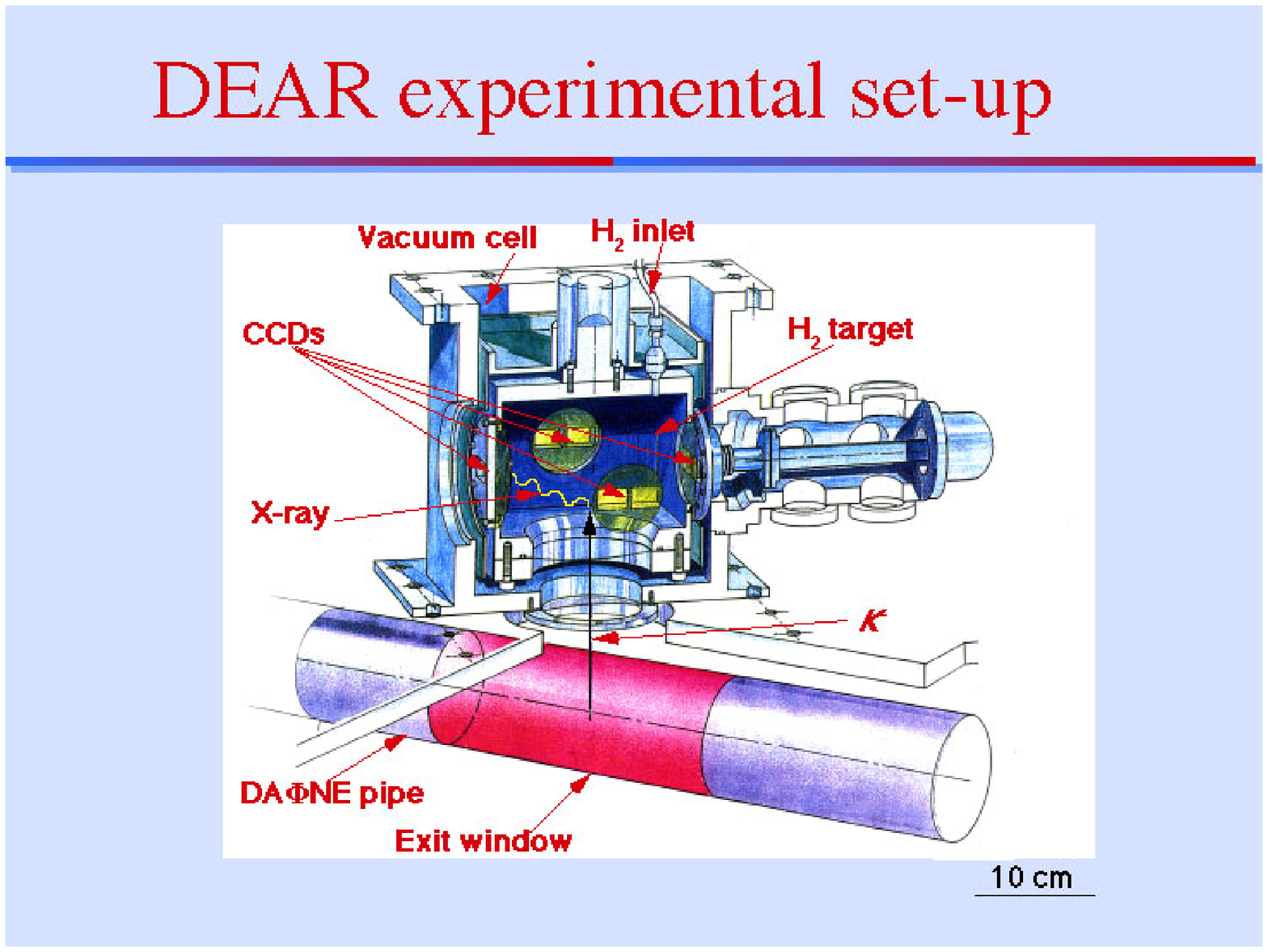}
\caption{Cut-out sketch of the DEAR setup.}
\label{deamerida96}
\end{figure}

The DEAR experiment \cite{guaraldo95} \cite{dearprop} 
will measure
the $KN$ scattering lengths by studying the shift and the width of the
energy levels of kaonic hydrogen and kaonic deuterium
atoms, i.e. atoms in which a $K^-$ from
the $\phi$ decay replaces the orbital electron in a hydrogen, or deuterium,
 target.
The negative kaon  then cascades,   initially through Auger
transitions, and
finally through emission of X-rays, whose energy is measured
by an array of Charge-Coupled Devices  (Fig.\ref{deamerida96}).
\par
\par
\begin{figure}[p]
\vspace{15cm}
\includegraphics{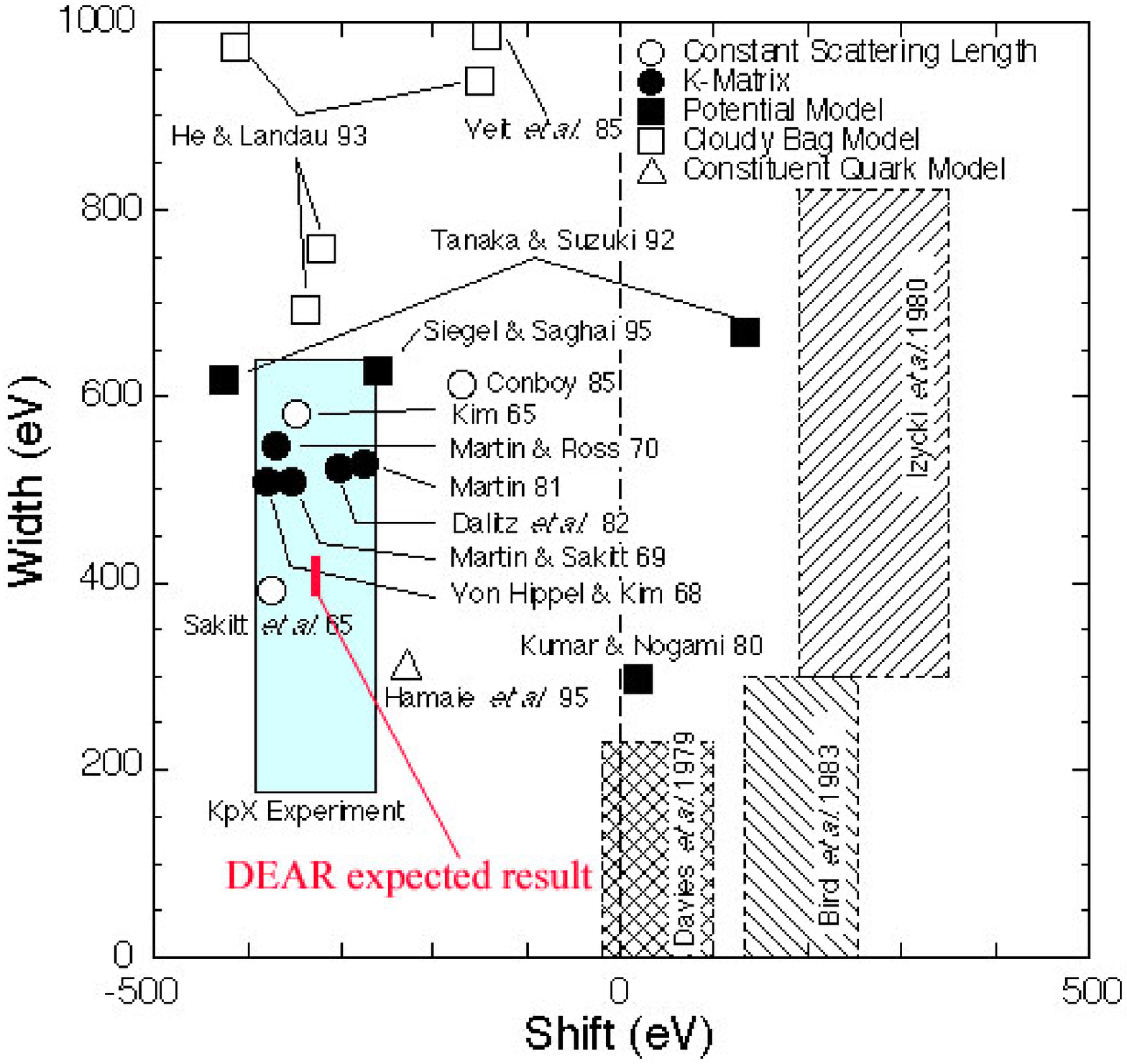}
\caption{Synopsys of theoretical predictions, experimental results,
and simulated DEAR results for shift $\Delta E$ and width $\Gamma$
of the $K_\alpha$ line of kaonic hydrogen.
}
\label{dearcomparison}
\end{figure}    
\par
The strong interaction between the captured $K^-$ and the target nucleus
shift the energy level of the lowest-lying atomic orbital
from their purely em level. Furthermore, the absorption process reduces
the  state lifetime, thus broadening the width of the X-ray transition.
\par
\par

\par
\begin{figure}[t]
\vspace{12cm}
\includegraphics{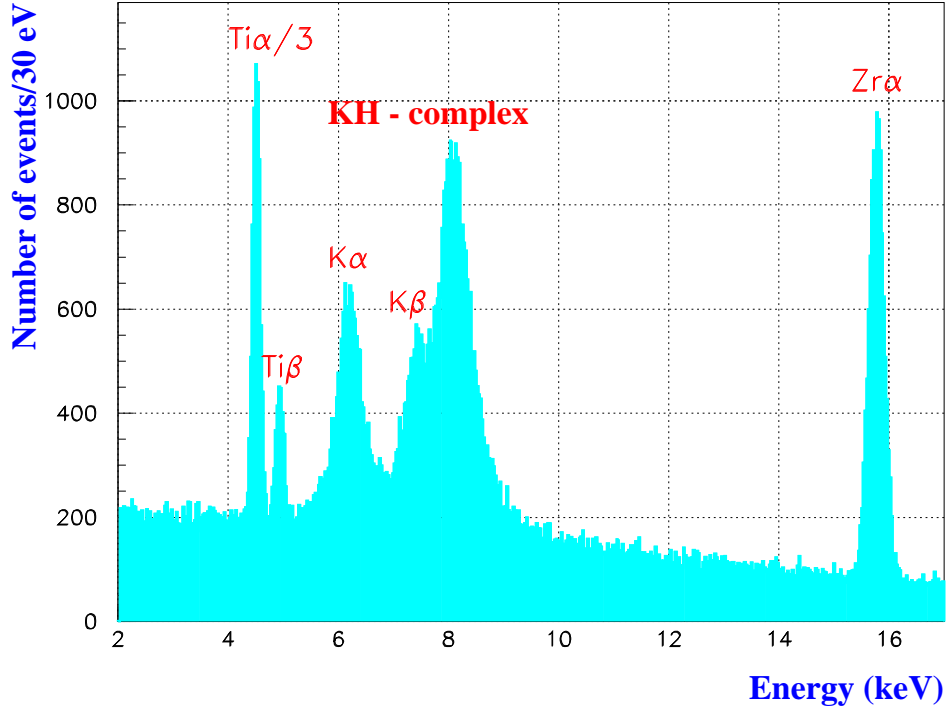}
\caption{
Simulated DEAR spectrum showing the $K_\alpha,K_\beta $ lines and
 the $K_{> \beta}$ complex, along with the $Ti$ and $Zr$ calibration
lines.
}
\label{dearpeaks}
\end{figure}    
\par
In the case of kaonic hydrogen, the  scattering 
length $a_{K^-p}$ is computed from the 
shift $\Delta E\equiv E_{exp}-E_{em}$
 of the ground level  from its computed em value, and the 
measured 
width $\Gamma$  of the X-ray line,  via the Deser formula
\begin{eqnarray}
   \Delta E + \frac{i\Gamma}{2} & = & c_{K^-p} a_{K^-p}     
\end{eqnarray}
where 
$ c_{K^-p} \equiv \alpha_S^3 \mu^2$, 
$\alpha_S$ is the strong-interaction fine-structure constant, and
$\mu$ the reduced mass.
The (unshifted) em energy levels are computed with a precision of about
1~eV, while  the experimental systematics in the detector is 
at the level of a few eV's.
\par
Experiments performed in the past are mutually inconsistent
and disagree with 
the analysis of low-energy scattering data,
 whilst
recent results from KEK \cite{iwasaki96} do confirm 
the extrapolation at threshold of scattering results
  (Fig.\ref{dearcomparison}).
\par
A  detailed simulation of signal and backgrounds, based
on a version of the GEANT3 code upgraded to treat photon energies
down to 1~keV \cite{petrascu97},
shows that both machine and physical background can be kept
under control (Fig.\ref{dearpeaks}) with the DEAR setup:
the prominent $K_\alpha$ peak shown (about 10,000 events)
can be accumulated with an integrated luminosity of about
$500\,{\rm pb}^{-1}$ (ref.\cite{guaraldo97}).
Superior energy resolution, and lesser background
with respect to the KEK experiment,
will dramatically reduce the error on the measurements
of both shift and width of the $K_\alpha$ line
in kaonic hydrogen, whilst the kaonic deuterium will be measured for the
first time
(Fig.\ref{dearcomparison}).
\par
In addition to the precise determination of low-energy parameters, 
it has been recently pointed out \cite{guaraldo97} how DEAR 
is able to determine with good precision
the so-called KN sigma term $(\sigma_{KN})$.
The sigma-term is a quantity which gives indication of the 
chiral-simmetry breaking terms in the strong Hamiltonian
\cite{kluge91}.
The theoretical evaluation of the sigma-term requires that the limit
of zero-mass pseudoscalar mesons is taken, and therefore 
approximations are required in order to use experimental data
obtained from physical particle processes \cite{diclaudio79},
\cite{gasser91}.
Theoretical studies suggest that $\sigma_{KN}$ terms
be much more sensitive to the $s\bar{s}$ content of the nucleon than
the pionic $\sigma_{\pi N}$ terms\cite{jaffe87}..
\par
The DEAR Collaboration (a total of 12 Institutes from 6 countries)
will take data as soon as the first stable beams circulate  in 
the \DAFNE rings.
\par

\section{Conclusions}
A  wealth of physics ranging from CP violation to hypernuclear studies
is expected from \DAFNE. Machine commissioning is in progress;
 construction of the detectors is well underway.
\par
I gratefully acknowledge the help and information given by my
colleagues on the \DAFNE project team, DEAR collaboration,
 FINUDA collaboration, and KLOE collaboration,
particularly G.~Vignola, C.~Guaraldo, C.~Petrascu, F.L.~Fabbri,
R.~Baldini, J.~Lee-Franzini, P.~Franzini, S.~Giovannella  and S.~Miscetti.
I also thank V.~Serdiouk for his comments on this manuscript.
Finally, I wish to thank the Organizing Committee  for a perfectly enjoyable
Symposium.
%

%
\end{document}